\documentclass{elsart}

\usepackage{epsfig}

\usepackage{amssymb}

\begin{document}

\begin{frontmatter}



\title{First Lasing of Volume FEL (VFEL) at Wavelength Range $\lambda \sim 4-6$ mm \thanksref{beltex}}
\thanks[beltex]{This work is carried out with financial support of private joint-stock company 
BelTechExport, Belarus}


\author{V. Baryshevsky, K. Batrakov, A. Gurinovich, I. Ilienko,} 
\author{A. Lobko,V.Moroz, P.Sofronov, V. Stolyarsky}

\address{22050, Bobruyskaya 11, Minsk, Belarus, Research Institute of Nucler Problem, Belorussian State University}

\begin{abstract}
First lasing of volume free electron laser (VFEL) is described. The
generating system consists of two metal diffraction grating with different
spatial periods. The first grating creates the conditions for Smith Purcell
emission mechanism. The second grating provides the distributed feedback for
emitted wave. The length of diffraction grating is 10 cm. Electron beam
pulse with a time duration $\tau \sim$ 10 ms has a sinusoidal form with
the amplitude varied from 1 to ~10 kV. The measured microwave
power reached the value of about 3-4 W in mm wavelength range. 
The generation stops at threshold current value.  
When the current tends to the threshold value, the region of generation tends
to a narrow band near to 5 kV. At higher current
values the radiation appears in electron energy range 5 - 7.5 KeV.
\end{abstract}


\end{frontmatter}

\section{Introduction}
\qquad The most essential feature of FEL and other types of generators is a
feedback, which is formed by a system of mirrors, or distributed feedback
based on diffraction in spatially periodic medium, when wave vectors of transmitted and
reflected waves are colinear. The distinction of volume
FEL (VFEL) is non-one-dimensional multi-wave volume
distributed feedback (VDFB). VDFB performs two basic functions
simultaneously:\\
1) it provides more effective interaction of an electromagnetic wave with
an electron beam due to new dispersion law; \\
2) it forms volume distributed mirror retaining radiation in
interaction region. 

\qquad 	It is well known that each radiative system is defined by its 
eigenmodes and by the so-called dispersion equation, 
which in the case of small perturbations (linear regime) describes possible 
types of waves in system and relation between frequency and wave number of 
system eigenmodes. Thorough analysis of properties of FEL dispersion 
equation is done in \cite{1_bvg}. It is shown there that \\
1. dispersion equation for FEL in collective regime coincides with that for 
conventional travelling wave tube amplifier (TWTA) \cite{2_bvg}; \\
2. FEL gain (increment of electron beam instability) 
is proportional to $\rho ^{1/3}$, where $\rho$ is the electron 
beam density.

\qquad For the first time the possibility of essential change of
instability of an electron beam moving in a spatially periodic medium was
indicated in  \cite{PhysRev}. In this paper the dispersion equations are 
obtained and investigated for conditions of multi-wave diffraction. It is
shown that there is a new law of electron beam instability in the points of 
diffraction roots degeneration. The amplification and generation gain
of electromagnetic wave are sharply changed in these points. In \cite{PhysRev}
this statement is derived from the fact that the instability increment in the point 
with s-times degeneration  is proportional to $\rho ^{1/(3+s)}$ 
(here $\rho $ is the electron beam density and it is supposed that the point in 
which s roots coincide is the point with (s-1)-times degeneration). 
This increment differs from the relevant increment for one-wave
system which is proportional to $\rho^{1/3}$. 
This result is also valid for electron beam which moves in vacuum close to the 
surface of spacially periodic medium \cite{Vacuum} (or
in a vacuum slot made inside a periodic medium). Explicit
expressions for dependence of starting current $j$ on interaction length $L$ are
obtained in degeneration points  \cite{bar1}: 
$j_{start}\sim 1/\{(kL)^{3}(k\chi _{\tau }L)^{2s}\}$. 
The advantages of VFEL are exhibited in wide spectral range from microwaves to X-rays
\cite{bar1,bar3,bar4}. In \cite{bar4} the experimental simulation of electrodynamic 
processes in the volume diffraction grating was performed for a millimeter wavelength
range. The possibility of obtaining of extremely high
Q-factor for a system with two strongly coupled waves is experimentally confirmed.

\qquad In the present work the first lasing of VFEL in millimeter wavelength range
is reported. The main parts of VFEL are two flat diffraction gratings with different spatial
periods. The first diffraction grating is used for radiation generation due to
Smith-Purcell radiating mechanism \cite{Smith}. The second grating provides
distributed feedback using Bragg dynamical diffraction
\cite{Czhang}, the grooves of diffraction grating are oriented at the 
nonzero angle to the direction of electron beam velocity. 
It should be noted, that generation in nonrelativistic 
TWT devices
in similar conditions is impossible at single harmonic, since the wavelength
of emitted radiation considerably exceeds period of TWT spiral (or corrugation period). 
Lasing of considered VFEL type for the first time
was described in \cite{Batr}, where the theoretical model of its operation was
presented. 
\section{Experimental setup and results}

\begin{figure}[h]
\epsfxsize = 14 cm 
\centerline{\epsfbox{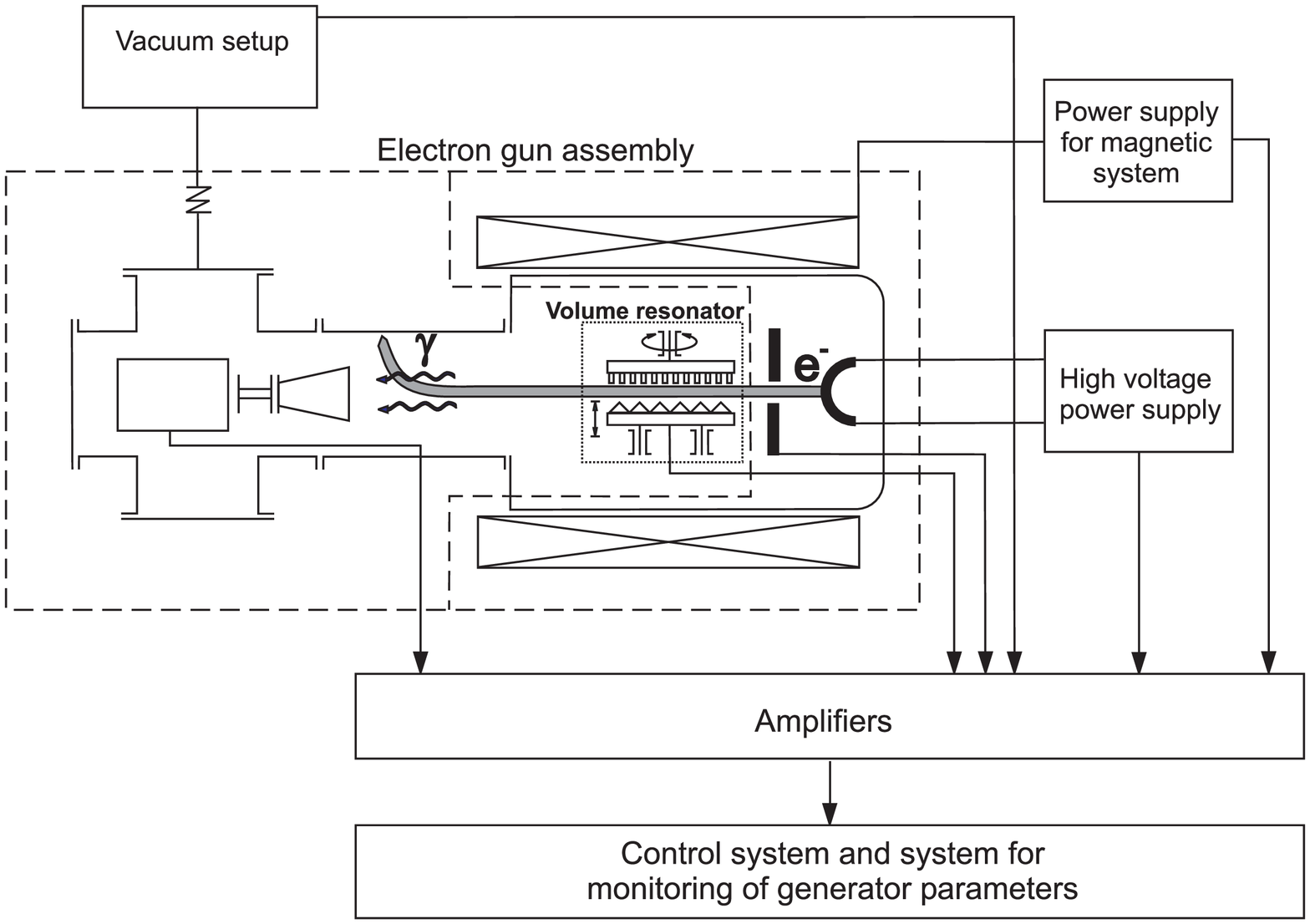}}
\caption{Block scheme of experimental setup for studing of VFEL generation.}
\label{block}
\end{figure}

\qquad Block-scheme of the experimental setup is illustrated in fig.\ref{block}.

\begin{figure}[h]
\epsfxsize = 7 cm 
\centerline{\epsfbox{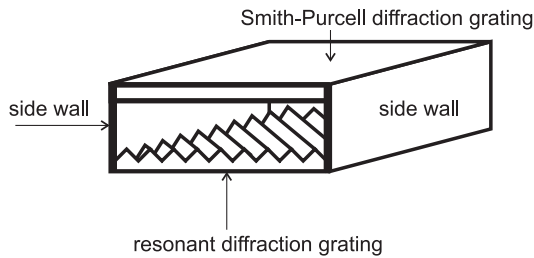}}
\caption{Block scheme of experimental setup for studing of VFEL generation.}
\label{resonator}
\end{figure}

\qquad Volume resonator is the basic functional part of VFEL (see Fig.\ref{resonator}). 
In considered setup
it is formed by two diffraction gratings with different periods and two smooth
side walls. Cross section of the resonator is rectangular and constant along
all its length. But the distance between diffraction gratings can be varied in different 
experiments. Radiation is output from the resonator through end-wall. 

\qquad The interaction of the first diffraction grating
(exciting grating) with the electron beam arouses generation of
Smith-Purcell radiation. The second (resonant) grating provides distributed
feedback of generated radiation with electron beam.
The distance between grooved surface of exciting grating and electron beam 
can be varied during experiment that allows to encrease efficiency of 
generation process. Rotation of resonant diffraction grating provides the 
possibility to change  radiation frequency by varying the orientation of grating 
grooves  with respect to electron beam velocity.
Mechanical tuning of the radiation
frequency is also provided by the adjustment of distance between gratings.

\qquad The sheet electron beam emitted by the thermal cathode (tungsten labilized
by barium or calcium iridate) is formed in cylindrical Peirce gun. It is 
guided in the VFEL resonator by the magnetic field $\sim 3 ~KGs$. 
Electrons are emitted in pulsed regime (unipolar pulse with
sinusoidal shape and pulse duration $\sim 10 ~ ms$) in
sequence of two or three voltage pulses (voltage amplitude can
vary from 1 to 10 KeV).

\qquad Generated radiation is output through a radiotransparent window
(plexiglas) to a radiation detector (thermistor detector M5-50 or
power meter M3-22A) with a bandwidth of detected radiation
54-78 GHz. 

\qquad Control and measurement system provides output of registed 
data to PC display. Also the following parameters are recorded on 
a hard disk drive:
\begin{itemize}
\item Cathode voltage;
\item Total electron beam current;
\item Current in a guiding magnetic system windings;
\item Grating current;
\item Power of microwave radiation.
\end{itemize}

\begin{figure}[h]
\epsfxsize = 11 cm 
\centerline{\epsfbox{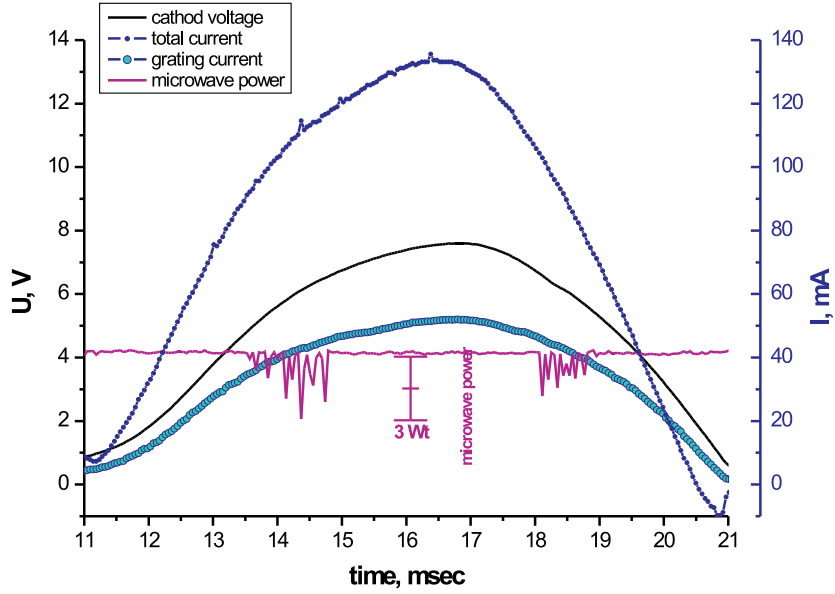}}
\caption{The oscillogram of VFEL lasing.}
\label{sep}
\end{figure}

\qquad The generation of microwave radiation
was detected at electron energy $\ge 5~ KeV$ in millimeter wavelength range ($\lambda \sim 4-6$ mm). 
Parameters of resonator are: 
the length of resonator is 100 mm;\\ 
the period of the exciting diffraction grating is 0.67 mm;\\ 
the period of the resonant diffraction grating is 3 mm. \\
The detected pulse power of generated radiation was about 3 - 4 W. Taking into account
the fact, that only the part of the electron beam with cross size 
$\delta \sim \frac{\lambda u}{4\pi c}<0.1~mm$  interacts effectively with 
electromagnetic wave the efficiency of
''working'' part of a beam can be evaluated as $\sim 10 \%$. The oscillogram 
of VFEL lasing is shown in fig. \ref{sep}.

\bigskip 

\begin{figure}[h]
\epsfxsize = 14 cm 
\centerline{\epsfbox{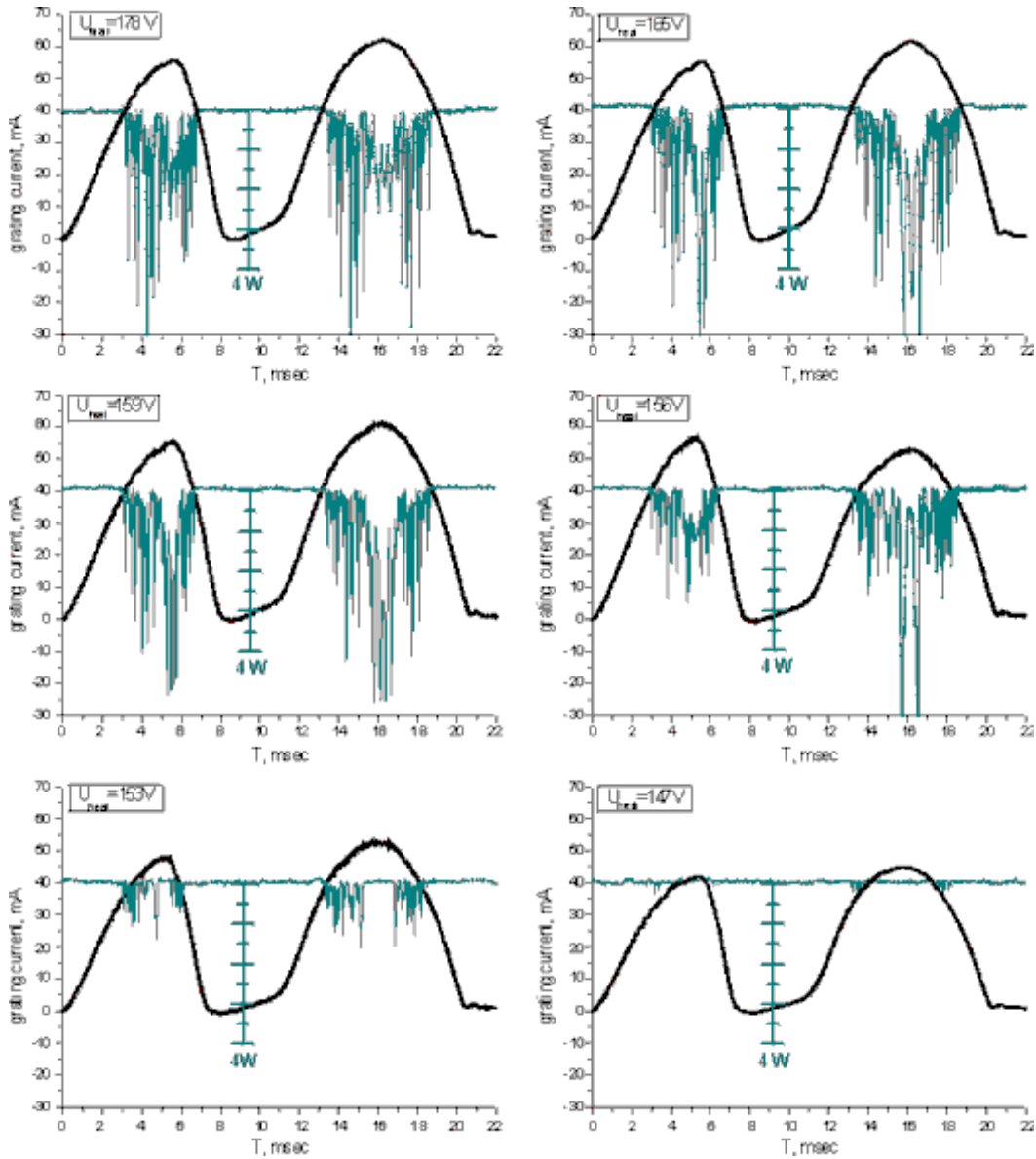}}
\caption{The oscillograms of dependence of VFEL lasing on the cathode heating voltage.}
\label{fil}
\end{figure}

\qquad The continuous generation of radiation was observed for electron beam currents 
higher than $ 35~mA$ . To define the threshold
conditions the beam current
was changed by the varying of the cathode heating voltage. 
Threshold conditions of generation are illustrated in fig.\ref{fil}, where
the dependence of grating current and emitted microwave radiation on 
cathode heating voltage is shown. The decreasing of heat voltage from 178 V to 158 V 
does not cause the decreasing of the grating current because of saturation mechanisn 
due to Child-Langmuir law. But at $U_{heat}\le 156 V$ the grating current decreases. 
It is clear from fig. \ref{fil} that reduction of the current leads to decreasing of radiation 
power. And the peaks on radiation power curve concentrates
near electron beam energy $\sim 5~keV$ when the grating current tends to the threshold value.
The peaks on radiation curve appears at minimum value of grating current  $\sim 35~ mA$.
The lasing starts from voltage $\sim 5~ kV$. This value is defined by the diffraction 
gratings period. As a result at the electron beam energy $\sim 5~ kV$ 
the mode with the highest Q-factor is excited. For this reason 
when the grating current tends to the threshold value, the region of generation tends
to a narrow band near to $ 5~ kV$. At higher current
values the radiation appears in electron energy range 5 - 7.5 KeV that demonstrates
the excitation of the other modes with smaller Q-factor, for which the
working current exceeds the starting that.  
 
\begin{thebibliography}{}


\bibitem{1_bvg} A.Gover, Z. Livni  {\em Optics Communications} {\bf 26} (1978), 375;
\bibitem{2_bvg} J.R.Pierci. {\em Travelling wave tubes} (Van Nostrand, Princeton 1950);
\bibitem{PhysRev} V.G.Baryshevsky, I.D.Feranchuk,  {\em Phys.Lett.\/} {\bf 102A} {1984} 141. 
\bibitem{Vacuum} V.G.Baryshevsky, {\em Doklady Akademii Nauk USSR\/} {\bf 299} {1988} 19.
\bibitem{bar1} V.G.Baryshevsky, K.G.Batrakov, I.Ya. Dubovskaya {\em Journ.Phys D.\/}{\bf 24} {1991}
1250.
\bibitem{bar3} V.G.Baryshevsky,  {\em NIM \/}{\bf 445A} {2000} 281.
\bibitem{bar4} V.G.Baryshevsky, K.G.Batrakov, I.Ya. Dubovskaya, V.A.Karpovich,
V.M.Rodionova, {\em NIM \/} {\bf 393A} {1997}  71-75.
\bibitem{Smith} Smith S.L., Purcell E.M., {\em Phys. Rev.\/} {1953} {\bf 91} 1069.
\bibitem{Czhang} Shih-Lin Chang, {\em Multiple Diffraction of X-Rays in Crystals\/} 
(Springer-Verlag, 1984).
\bibitem{Batr} V.G.Baryshevsky, K.G.Batrakov, V.I.Stolyarsky {\  Proc. of the 21
International FEL Conference \/} {1999} II-37. 
\end{thebibliography}
\end{document}